\documentclass[aps,pre,reprint,amsmath,amssymb,superscriptaddress,longbibliography,floatfix,]{revtex4-2}

\usepackage{graphicx}
\usepackage{bm}

\begin{document}

\title{Network versus market information in coupled opinion–return dynamics}
\author{Yue Yu}
\email{y.y@ugent.be}
\affiliation{Department of Economics, Ghent University, Ghent, Belgium}
\author{Luis E C Rocha}
\email{luis.rocha@ugent.be}
\affiliation{Department of Economics, Ghent University, Ghent, Belgium}
\affiliation{Department of Physics and Astronomy, Ghent University, Ghent, Belgium}

\begin{abstract}
When making investment decisions, investors can use common signals from the market and local signals from their social networks. We develop an agent-based opinion-return model in which agents follow fixed trading strategies and respond to either local network information or a common market signal. In our model, trend followers and contrarians select either a local network signal or a common market signal and then update their buy, hold, or sell opinions by following or opposing the signal, respectively. The aggregate opinion imbalance determines the return, which sets the market signal. We find that three collective regimes emerge depending on how much network information is used and the fraction of contrarians in the population. Persistent directional states occur when trend followers dominate, market-wide reversals appear when a contrarian-rich population relies on the common market signal, and opposing opinions coexist with small aggregate returns when the same population relies on network signals. With a common market signal, the majority strategy has the higher payoff scores, whereas increasing reliance on network signals reduces this difference. When only network information is used, higher network clustering is associated with greater polarisation and smaller return magnitudes when trend followers dominate. Degree heterogeneity has little effect on aggregate measures but changes the payoff scores of high- and low-degree agents. In contrarian-rich populations, local information thus reduces aggregate return fluctuations even as opposing opinions persist.
\end{abstract}

\maketitle

\section{Introduction}

Financial markets are complex systems composed of heterogeneous investors whose interactions jointly shape asset-price dynamics~\cite{lux1999}. Investors do not make decisions in isolation. They respond to changes in market conditions while also observing and communicating with other market participants~\cite{banerjee1992,hong2005thy}. Their resulting opinions may be interpreted as intentions to buy, hold, or sell an asset. When aggregated, these individual opinions generate excess demand and affect market prices. Price changes, in turn, provide new information that influences subsequent investment decisions. Investor opinions and market prices therefore form a coupled system in which individual behaviour produces aggregate market signals that feed back into future individual behaviour~\cite{hommes2005coordination,veldkamp2006information}. The feedback therefore depends on both the information source and the response rule.

Opinion-dynamics models provide a theoretical framework for studying how local interactions generate collective behaviour. In the voter model, individuals update their opinions by imitating neighbouring agents, whereas majority-rule models assume that individuals adopt the prevailing opinion within a local group~\cite{HolleyLiggett1975,Galam2002}. More recent models incorporate heterogeneous influence, individual persistence, social environments, and adaptive connections. Ojer et al.~\cite{ojer2025}, for example, combined social learning with individuals' preferences for maintaining their initial opinions and showed that network heterogeneity and hub nodes can alter transitions between consensus and polarisation. The influence of neighbours' wider opinion environments has also been considered~\cite{krishnagopal2025}. Agents at different network positions can therefore receive different local signals. Network connectivity and position also affect information aggregation in models of social learning~\cite{DeMarzoVayanosZwiebel2003,GolubJackson2010}. Filtering which neighbouring opinions agents encounter can further change collective opinion distributions, even without modifying the underlying social network~\cite{PerraRocha2019}.

In agent-based financial markets, correlated trading decisions can generate large fluctuations in aggregate demand~\cite{cont2000}. Local interactions, investor heterogeneity, and trading frictions can also produce irregular price dynamics~\cite{iori2002}. Spin-based and cellular-automaton models have been used to study the relation between local trading activity and aggregate price fluctuations ~\cite{SznajdWeronWeron2002,KaizojiEtAl2002,BartolozziThomas2004}. Other models explicitly connect evolving beliefs to trading behaviour. Restocchi et al.~\cite{restocchi2023} used a bounded-confidence process in which sufficiently similar investors adjust their beliefs toward one another, with the resulting beliefs determining trades and market prices. Di Francesco and Torren-Peraire~\cite{difrancesco2025information} considered a market in which some investors receive direct signals about fundamentals, while others obtain information through their network neighbours. Empirical financial returns, however, show statistical regularities beyond those captured by such minimal models~\cite{Cont2001StylizedFacts}.

In majority-opinion models combining local peer influence with an external signal, network clustering can reinforce the external influence, whereas degree heterogeneity can weaken it~\cite{VanSanten2024}. Financial opinion models further distinguish between local and global information. Local majority alignment and global contrarian influence have been combined in majority-vote market models~\cite{VilelaEtAl2019}. In a three-state model, noise traders follow the local majority, while contrarians respond to the global market state by taking the minority position~\cite{zubillaga2022}. The effects of network topology and trader composition on these dynamics have been examined using scale-free, small-world, and random networks~\cite{zubillaga2025}. The results indicate that connectivity, trader composition, and behavioural parameters can be more influential than network type alone. A related financial-market opinion model by Granha et al. \cite{Granha2022PNAS} also couples neighbourhood-based responses of one investor category to market-based responses of another. Other market models combine public, private, and network information, or examine information flows and imitation separately \cite{HarrasSornette2011,Biondo2020}. In the three-state local/global models just described, however, information exposure is prescribed by trader type. This makes it difficult to separate the effect of the signal from the response to it.

Price dynamics also depend on the composition of trading strategies. Heterogeneous expectations can generate instability through nonlinear price feedback~\cite{brock1998}, while simple trading rules combined with price impact can produce endogenous price fluctuations~\cite{FarmerJoshi2002}. Trend-following and contrarian interactions have also been shown to produce different dynamical regimes \cite{GiardinaBouchaud2003}. Bornholdt~\cite{bornholdt2001} combined local imitation with a global incentive to act against the market majority, producing intermittent fluctuations and expectation-driven price movements. Trend-following, contrarian, and fundamental strategies have subsequently been incorporated into models in which investors respond differently to recent price trends, deviations from fundamental value, or past strategy performance~\cite{schmitt2021,anufriev2013,difrancesco2025sentiment,gardini2025}.

We separate information source from response strategy in a three-state opinion–return model. In the proposed model, both trend followers and contrarians select between a common market signal and local social network signals, while the composition of behavioural strategies is controlled separately. The aggregate opinion imbalance determines the return, which feeds back into the common market signal. By varying the probability of selecting network information and the fraction of contrarians, we distinguish changes associated with information source from those associated with strategy composition. Network position has also been associated with differences in trading outcomes~\cite{OzsoylevEtAl2014}. We then compare Watts-Strogatz and configuration networks, examining separately the changes in collective observables and the payoff differences between network positions.

\section{Methods}
\label{methods}

\subsection{Model}
\begin{figure}
    \centering
    \includegraphics[width=\columnwidth]{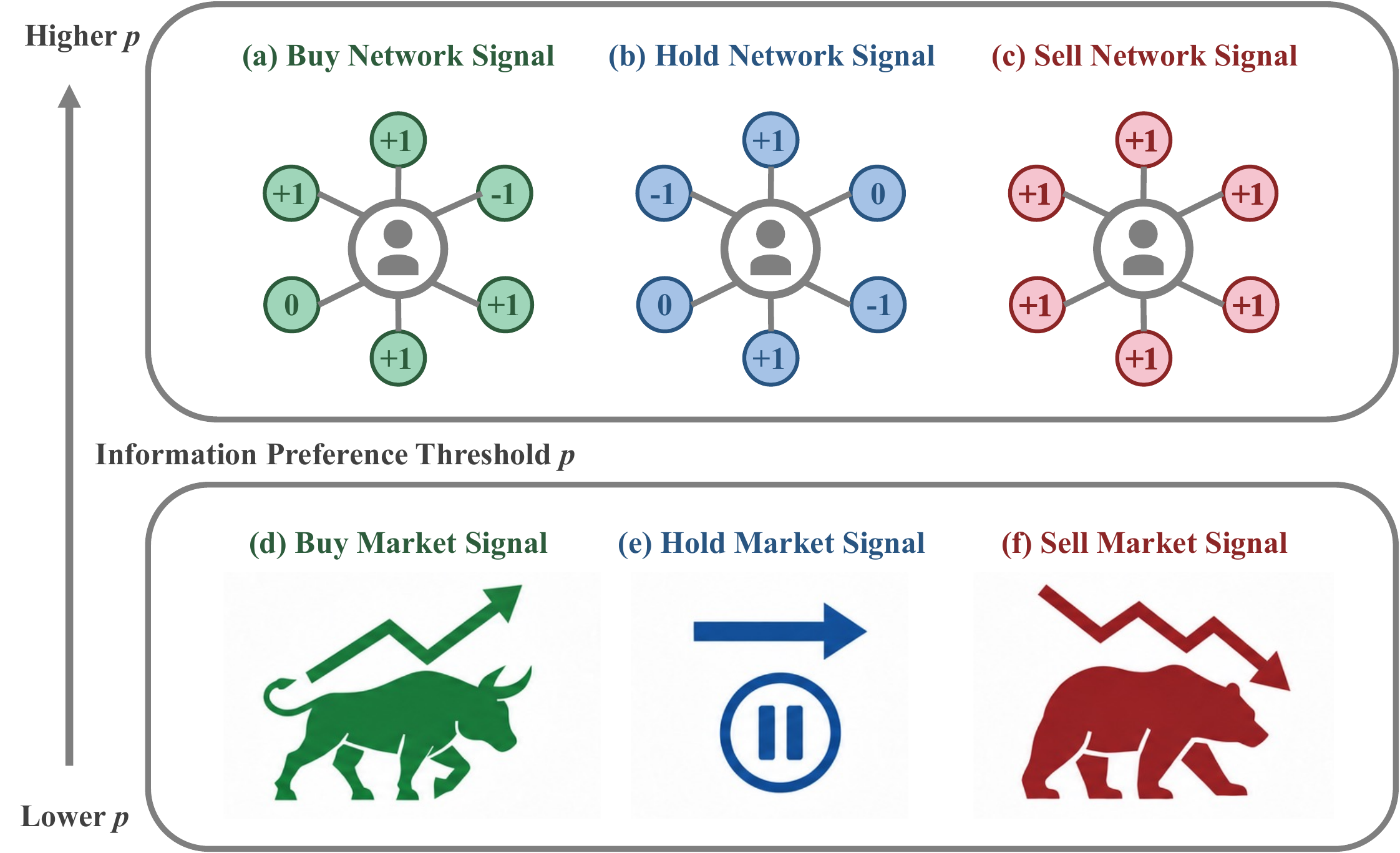}
    \caption{Schematic representation of the coupled opinion-return dynamics.}
    \label{fig:framework}
\end{figure}

We develop an agent-based model with local network signals, a common market signal, and heterogeneous trading strategies. Each agent $i$ holds an opinion $o_i(t)\in\{-1,0,1\}$, representing sell, hold, or buy intentions at time $t$, respectively (Fig.~\ref{fig:framework}).

Each agent observes the opinions of its network neighbours. We represent these interactions by a network of $N$ investors, in which nodes denote investors and edges represent social connections~\cite{Newman2018}. The set of neighbours of agent $i$ is denoted by $\mathcal{N}_i$.

At time $t$, the local information available to agent $i$ is characterised by the average opinion of its neighbours,
\begin{equation}
\overline{O}_i(t)
=
\frac{1}{|\mathcal{N}_i|}
\sum_{j\in\mathcal{N}_i}o_j(t).
\end{equation}
The corresponding network signal is defined as
\begin{equation}
s_i^{\mathrm{net}}(t)
=
\operatorname{sign}\!\left[\overline{O}_i(t)\right]
=
\begin{cases}
 1,  & \overline{O}_i(t)>0,\\
 0,  & \overline{O}_i(t)=0,\\
-1,  & \overline{O}_i(t)<0.
\end{cases}
\end{equation}
A positive (negative) network signal indicates that the local opinion environment favours buying (selling) the asset, whereas a zero signal indicates zero net opinion in the neighbourhood.

In contrast to the network signal, which depends on an agent's local neighbourhood, the market signal is common to all agents and is derived from the most recent log-return. The log-return at time $t$ is defined as
\begin{equation}
r(t) = \log P(t)-\log P(t-1),
\end{equation}
where $P(t)$ denotes the asset price at time $t$.

The market signal is obtained by comparing the scaled return $\alpha r(t)$ with a threshold $\theta$,
\begin{equation}
s^{\mathrm{mkt}}(t)
=
\begin{cases}
 1,  & \alpha r(t)>\theta,\\
 0,  & |\alpha r(t)|\leq\theta,\\
-1,  & \alpha r(t)<-\theta,
\end{cases}
\end{equation}
where $\alpha>0$ rescales the return before thresholding, and $\theta\geq0$ defines a threshold for the market signal. Sufficiently positive or negative returns generate buying or selling signals, respectively. When $|\alpha r(t)|\leq\theta$, the market signal is zero.

Agents are divided into three types: trend followers, contrarians, and noise traders. Their fractions are denoted by $f_{\mathrm{T}}$, $f_{\mathrm{C}}$, and $f_{\mathrm{N}}$, respectively, with
\begin{equation}
f_{\mathrm{T}}+f_{\mathrm{C}}+f_{\mathrm{N}}=1.
\end{equation}

Trend followers and contrarians select between network and market information before updating their opinions. We use a parameter $p$ to set the probability of selecting network information. At each update, agent $i$ draws a random variable $u_i(t)$ from a uniform distribution. The selected information signal is
\begin{equation}
I_i(t) =
\begin{cases}
s_i^{\mathrm{net}}(t), & u_i(t)<p,\\
s^{\mathrm{mkt}}(t),   & u_i(t)\geq p,
\end{cases}
\end{equation}
where $u_i(t)\sim\mathcal{U}(0,1)$. Thus, $p=0$ corresponds to exclusive reliance on market information, whereas $p=1$ corresponds to exclusive reliance on network information. These two limits isolate the feedback channels. At $p=0$, the network does not enter the opinion-update
rule; at $p=1$, market returns are generated by the opinion dynamics but are not fed back into opinion updating.

At each time step, all agents update their opinions using the configuration at time $t$. The opinion-update rule is
\begin{equation}
\label{eq:opinion_update}
o_i(t+1)
=
\begin{cases}
\operatorname{sign}\!\left[I_i(t)\right],
& \text{trend follower},\\
-\operatorname{sign}\!\left[I_i(t)\right],
& \text{contrarian},\\
\xi_i(t),
& \text{noise trader},
\end{cases}
\end{equation}
where $\xi_i(t)$ is drawn independently from 
$\{-1,0,1\}$ with equal probability.

After all agents have updated their opinions, their individual trading intentions are aggregated into excess demand,
\begin{equation}
D(t+1) = \sum_{i=1}^{N}o_i(t+1).
\end{equation}

The log-price evolves according to the normalised excess demand~\cite{FarmerJoshi2002},
\begin{equation}
\log P(t+1) = \log P(t)+ \lambda\frac{D(t+1)}{N},
\end{equation}
where $\lambda>0$ is the price-impact parameter. Combining this price-update rule with the definition of the log-return gives
\begin{equation}
r(t+1) = \lambda\frac{D(t+1)}{N}.
\label{eq:return_from_demand}
\end{equation}

In this stylised price-impact rule, the sum of current directional opinions is used as an excess-demand proxy. It does not distinguish new orders from changes in pre-existing asset holdings. Accordingly, a persistent nonzero opinion imbalance generates a persistent nonzero return.

Using $o_i(t)$ as a signed directional exposure, we define the payoff scores

\begin{equation}
\label{eq:payoff}
\pi_i(t) = o_i(t)r(t) - c\left|o_i(t)-o_i(t-1)\right|,
\end{equation}

where $c\geq 0$ is the cost per unit change in opinion. Because the same opinions also contribute to the contemporaneous price movement through Eq.~\eqref{eq:return_from_demand}, and the model does not explicitly track asset holdings or cash balances, this score is not a realised, self-financing trading profit.

\subsection{Observables}

Cross-sectional opinion polarisation is defined as
\begin{equation}
\label{eq:polarization}
\mathcal P(t)=\left\langle o_i^2(t)\right\rangle-\left\langle o_i(t)\right\rangle^2,
\end{equation}
where $\langle\cdot\rangle$ denotes an average across investors. Let $\overline{\cdot}$ denote a temporal average over the post-burn-in observations. The polarisation is defined as the time-averaged cross-sectional opinion variance $\mathcal P_I=\overline{\mathcal P(t)}$. It is zero when all investors hold the same opinion and increases as buy and sell opinions coexist within the population.

Aggregate opinion imbalance is defined as
\begin{equation}
\label{eq:imbalance}
I(t) = f_{\mathrm B}(t) - f_{\mathrm S}(t),
\end{equation}
where $f_{\mathrm B}(t)$ and $f_{\mathrm S}(t)$ are the fractions of buy and sell opinions at time $t$. The hold fraction is $f_{\mathrm H}(t)=1-f_{\mathrm B}(t)-f_{\mathrm S}(t)$. The reported mean absolute imbalance, $|I|=\overline{|I(t)|}$, measures the strength of directional dominance, while the reported mean hold fraction, $f_{\mathrm H}=\overline{f_{\mathrm H}(t)}$, gives the time-averaged fraction of agents in the hold state.

The switching intensity is defined as
\begin{equation}
\mathcal S_I=\frac{1}{T'-1}\sum_{t=1}^{T'-1}\mathbf 1\!\left[I(t)I(t+1)<0\right],
\end{equation}
where $T'$ is the number of retained observations and $\mathbf 1[\cdot]$ is the indicator function. It measures the frequency of direct reversals between buy- and sell-dominated states; transitions through zero are not counted.

The absolute mean return is $\mathcal D_r=|\overline r|$, which measures persistent directional return movement. The mean absolute return magnitude is $\mathcal M_r=\overline{|r|}$, which measures the typical size of returns irrespective of direction. Return volatility is defined as $\sigma_r=\sqrt{\overline{(r-\overline r)^2}}$, while large return magnitudes are summarised by the 95th percentile of absolute returns, $\mathcal T_{0.95}=Q_{0.95}(|r|)$.

The bimodality coefficient~\cite{pfister2013good} is defined as
\begin{equation}
\mathcal B_r=\frac{g_r^2+1}{\kappa_r+3(T'-1)^2/[(T'-2)(T'-3)]},
\end{equation}
where $g_r$ and $\kappa_r$ are the sample skewness and excess kurtosis of returns. Values above the reference value $5/9$ are commonly used as an indication of bimodality, although skewed unimodal distributions can also give high values.

Let $e_{ab}(t)$ denote the fraction of directed edges from opinion state $a$ to state $b$, counting each undirected edge in both directions, where $a,b\in\{-1,0,+1\}$, and let $q_a(t)=\sum_b e_{ab}(t)$. Opinion assortativity~\cite{newman2003mixing} is defined as
\begin{equation}
\mathcal A_o(t)=\frac{\sum_a e_{aa}(t)-\sum_a q_a^2(t)}{1-\sum_a q_a^2(t)}.
\end{equation}
The reported value is $\mathcal A_o=\overline{\mathcal A_o(t)}$. Positive values indicate locally homogeneous opinion clusters, values near zero indicate approximately random mixing, and negative values indicate a tendency for different opinions to be connected.

\begin{figure*}[tb]
    \centering
    \includegraphics[width=\textwidth]{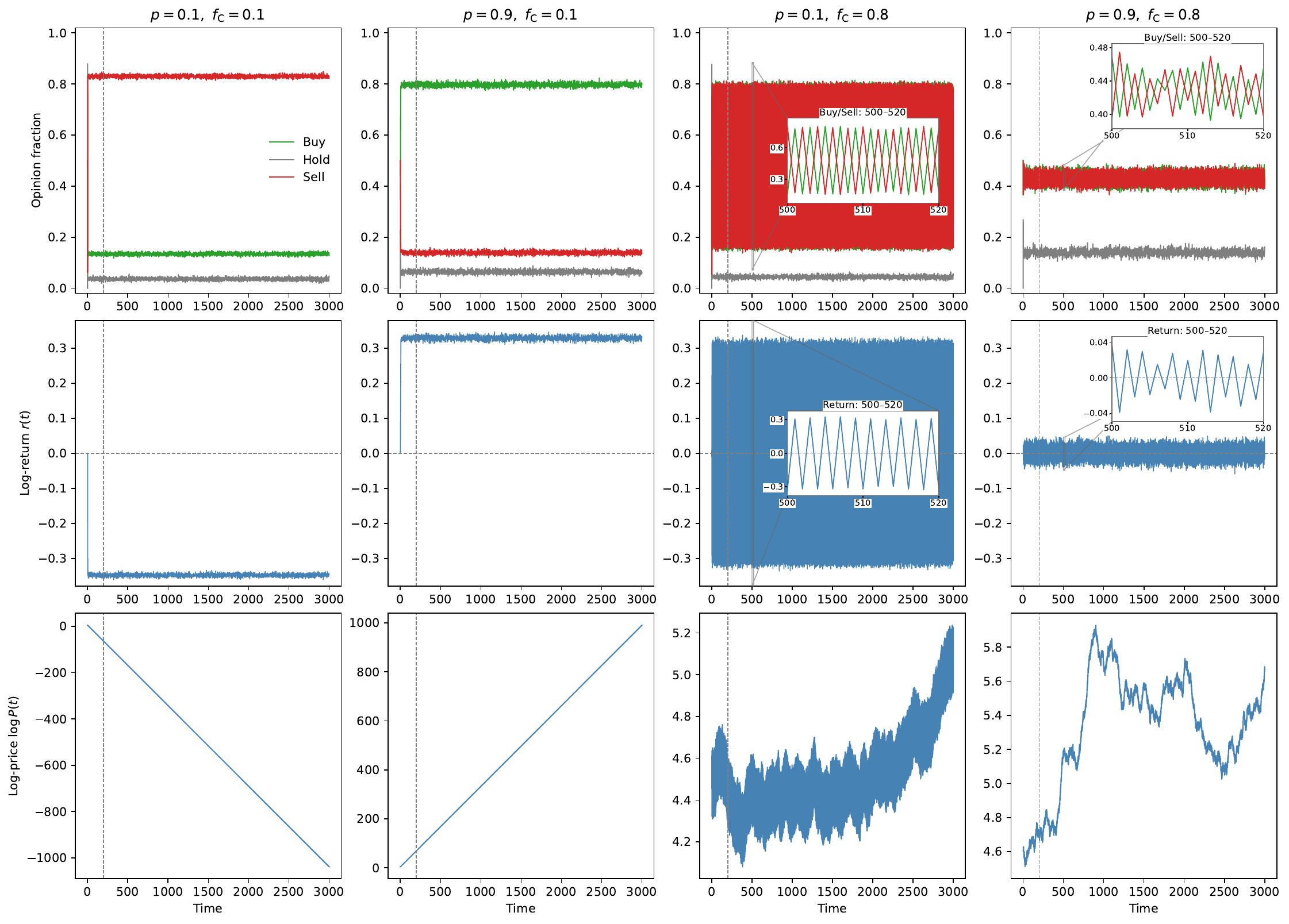}
    \caption{Representative time series of opinion fractions, log returns, and log prices for four combinations of the probability of selecting network information $p$ and the contrarian fraction $f_{\mathrm C}$. From left to right, the columns correspond to $(p,f_{\mathrm{C}})=(0.1,0.1)$, $(0.9,0.1)$, $(0.1,0.8)$, and $(0.9,0.8)$. Each column shows one realisation. The rows display the fractions of buy, hold, and sell opinions (top), the log return $r(t)$ (middle), and the log price $\log P(t)$ (bottom). The vertical dashed line marks the end of the burn-in period, and the horizontal dashed line in the middle row indicates zero return.}
    \label{fig:representative_regimes}
\end{figure*}

\subsection{Simulation setup}

The baseline analysis is conducted on Erdős-Rényi (ER) networks with $N=1000$ investors and expected mean degree $\langle k\rangle=6$~\cite{Newman2018}. Each realisation is simulated for $T=3000$ time steps, with the first $200$ observations discarded as burn-in. Buy and sell opinions are initially equally represented and randomly distributed across the network. Agent types are assigned at the beginning of each simulation and remain fixed throughout the dynamics. We set the noise-trader fraction to $f_{\mathrm N}=0.1$ and use $\lambda=0.5$, $\alpha=1$, $\theta=0$, and $c=0.01$ as the baseline parameter values. Because $\theta=0$ and $\alpha>0$, the sign of a nonzero market return, and hence the baseline market signal, does not depend on the numerical value of $\alpha$. The information-preference parameter $p$ and contrarian fraction $f_{\mathrm C}$ are varied systematically, while the trend-follower fraction is determined by $f_{\mathrm T}=1-f_{\mathrm C}-f_{\mathrm N}$.

To study the effects of network structure, we additionally consider Watts-Strogatz (WS) and configuration-model networks while holding network size and mean degree approximately constant~\cite{Newman2018}. Clustering in the WS networks is varied through the rewiring probability, whereas degree heterogeneity in the configuration networks is varied through the degree-distribution exponent. The node-position analysis compares investors in the top and bottom quartiles of local clustering in WS networks and node degree in configuration networks. We also repeat the strategy-payoff analysis under alternative switching-cost assumptions to assess the robustness of the results.

For each parameter combination, we generate 10 independent network realisations and conduct 10 independent dynamical runs on each network, giving 100 runs per parameter combination. Repeated runs are first averaged within each network. Reported ensemble means and approximate 95\% confidence intervals are then calculated across the 10 independent network realisations, with confidence-interval half-width $1.96s/\sqrt{10}$, where $s$ is the sample standard deviation of the ten network-level means.

\begin{figure*}[tb]
    \centering
    \includegraphics[width=\textwidth]{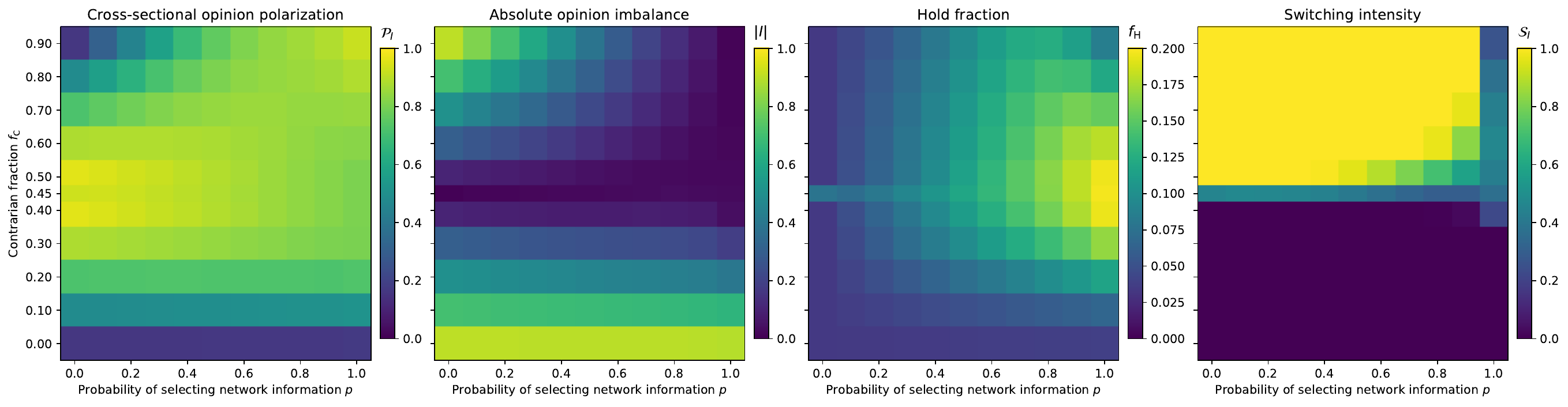}
    \caption{Collective opinion dynamics as functions of the probability of selecting network information $p$ and the contrarian fraction $f_{\mathrm{C}}$. From left to right, the panels show cross-sectional opinion polarisation $\mathcal{P}_I$, absolute opinion imbalance $|I|$, hold fraction $f_{\mathrm H}$, and switching intensity $\mathcal{S}_I$. A switch is counted when $I(t)I(t+1)<0$. Each measure is calculated over the post-burn-in period and averaged across independent simulation runs.}
    \label{fig:opinion_maps}
\end{figure*}

\section{Results}
\label{results}

\subsection{Representative regimes}

Figure~\ref{fig:representative_regimes} presents representative realisations for four combinations of the probability of selecting network information $p$ and the contrarian fraction $f_{\mathrm{C}}$. When the contrarian fraction is low, $f_{\mathrm{C}}=0.1$, both realisations develop a persistent directional state. At $p=0.1$, sell opinions remain dominant, producing persistently negative returns and a downward price trend. At $p=0.9$, buy opinions instead remain dominant, producing persistently positive returns and an upward price trend. The update rules are symmetric under the exchange of buy and sell opinions. The opposite price trends therefore reflect the direction selected in each realisation, rather than a systematic directional effect of $p$. Once selected, that direction is reinforced by the trend-follower majority through positive feedback, regardless of whether investors rely predominantly on market or neighbourhood information.

At $f_{\mathrm{C}}=0.8$ and $p=0.1$, investors rely mainly on the common market signal. Because most investors respond contrarily to the same signal, their coordinated actions repeatedly reverse the aggregate opinion imbalance. The system consequently switches between buy- and sell-dominated states, generating large positive and negative returns and a non-monotonic log-price trajectory. At $f_{\mathrm{C}}=0.8$ and $p=0.9$, investors rely primarily on local network signals. Their responses are therefore less synchronised, allowing opposing buy and sell opinions to coexist and partially offset one another at the aggregate level. The opinion imbalance remains closer to zero, return fluctuations are substantially smaller, and the log price varies on a much smaller scale than in the trend-follower-dominated cases. Since $r(t)=\lambda I(t)$, residual opinion imbalances still produce changes in log price.

The representative trajectories illustrate three dynamical regimes: persistent directional states when trend followers dominate, oscillatory market-wide reversals when a contrarian-rich population relies on the common market signal, and coexistence of opposing opinions with small aggregate returns when the same population relies on network signals. Appendix~\ref{app:representative_regimes} compares imbalance distributions and cross-sectional opinion fractions, distinguishing temporal reversals from the coexistence of buy and sell opinions at a given time. The return distributions also show both signs across independent runs at low $f_{\mathrm C}$, consistent with the buy-sell symmetry of the model.

\subsection{Collective dynamics}

Figure~\ref{fig:opinion_maps} shows the polarisation, absolute opinion imbalance, hold fraction, and switching intensity as functions of $p$ and $f_{\mathrm C}$. The dependence on $p$ changes with the contrarian fraction. At $p=0$ and $\theta=0$, trend followers and contrarians respond to the same market signal. Their contribution to the excess demand can then be written explicitly. Eqs.~\eqref{eq:opinion_update} and \eqref{eq:return_from_demand} give

\begin{equation}
\label{eq:common_information_benchmark}
r(t+1)=\lambda(1-f_{\mathrm N}-2f_{\mathrm C})\operatorname{sgn}[r(t)]+\eta(t+1)
\end{equation}

where $\mathcal{V}_{\mathrm N}$ is the set of noise traders
and $\eta(t+1)=(\lambda/N)\sum_{i\in\mathcal{V}_{\mathrm N}}\xi_i(t)$ represents their contribution. For independent, uniformly distributed three-state noise, ${\mathbb{E}[\eta]}=0$ and $\operatorname{Var}(\eta)=2\lambda^2 f_{\mathrm N}/(3N)$. The deterministic coefficient changes sign at

\begin{equation}
\label{eq:contrarian_threshold}
f_{\mathrm C}^* = \frac{1-f_{\mathrm N}}{2} = 0.45\qquad (f_{\mathrm N}=0.1).
\end{equation}

For $f_{\mathrm C}<f_{\mathrm C}^*$, the return sign persists in the deterministic limit, whereas for $f_{\mathrm C}>f_{\mathrm C}^*$, it alternates. The noise contribution decreases as $N^{-1/2}$. For $p>0$, the neighbourhood signals enter the update rule and this relation no longer applies.

\begin{figure*}[tb]
    \centering
    \includegraphics[width=\textwidth]{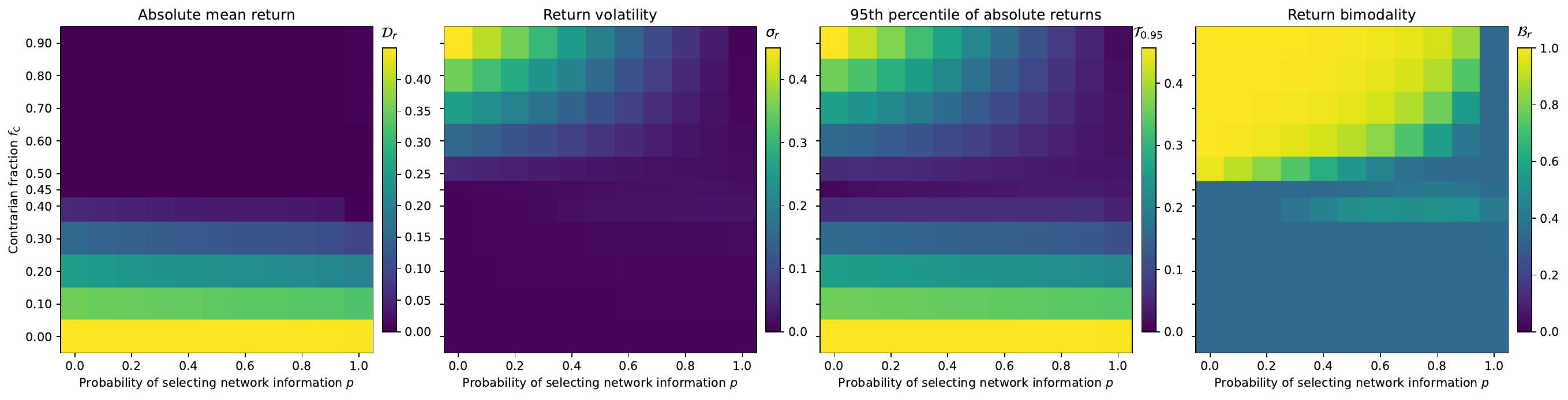}
    \caption{Market dynamics as functions of the probability of selecting network information $p$ and the contrarian fraction $f_{\mathrm{C}}$. From left to right, the panels show the absolute mean return $\mathcal{D}_r$, return volatility $\sigma_r$, the 95th percentile of absolute returns $\mathcal{T}_{0.95}=Q_{0.95}(|r|)$, and return bimodality $\mathcal{B}_r$. Each measure is calculated over the post-burn-in period and averaged across independent simulation runs.}
    \label{fig:market_maps}
\end{figure*}

When the contrarian fraction is low, $p$ has little effect on the opinion dynamics. The mean absolute imbalance remains high and the switching intensity remains close to zero across the full range of $p$. Because trend followers constitute the majority, they reinforce whichever directional signal becomes dominant after the initial stochastic fluctuations. Once a persistent directional state has formed, the market signal and most neighbourhood signals tend to point in the same direction. At intermediate contrarian fractions, the absolute imbalance is substantially smaller because trend followers and contrarians respond oppositely to the same information signal. When investors rely mainly on common market information, the two strategies form groups with opposing opinions, giving relatively high polarisation but a smaller aggregate imbalance. As $p$ increases, heterogeneous neighbourhood signals weaken this coordination. The hold fraction also increases as more agents respond to zero neighbourhood signals. Cross-sectional polarisation consequently decreases as reliance on network information increases.

When the contrarian fraction is high, the dependence on $p$ reverses. At low $p$, most investors respond contrarily to the same market signal and therefore act in a coordinated manner. The population has a strong directional majority at each time, resulting in comparatively low cross-sectional polarisation and a large absolute imbalance. However, the contrarian response reverses the sign of the aggregate state, so the directional majority switches frequently and $\mathcal S_I$ is high. As $p$ increases, investors receive heterogeneous network signals from their local neighbourhoods. Their actions become less synchronised, reducing both the aggregate imbalance and the frequency of market-wide reversals. The smaller aggregate imbalance does not imply consensus since opposing opinions increasingly coexist, and cross-sectional polarisation rises.

Because the log return is proportional to the aggregate opinion imbalance, $r(t)=\lambda I(t)$, these opinion regimes translate directly into distinct market outcomes. Figure~\ref{fig:market_maps} maps the absolute mean return $\mathcal{D}_r=|\overline r|$, return volatility $\sigma_r$, the high absolute-return quantile $\mathcal{T}_{0.95}$, and return bimodality $\mathcal{B}_r$ across the same parameter space.

At low contrarian fractions, the absolute mean return is high, whereas within-run return bimodality is comparatively low. Each realisation therefore tends to remain concentrated around either a positive or a negative return mode rather than repeatedly switching between them, as the trend-follower majority reinforces the selected direction. At intermediate contrarian fractions, the absolute mean return and high absolute-return quantile decline because the opposing responses of trend followers and contrarians reduce the aggregate opinion imbalance. At high contrarian fractions, information preference has a pronounced effect on market dynamics. When $p$ is low, volatility, high absolute-return quantile, and within-run bimodality are high, whereas the absolute mean return remains close to zero. Thus, large positive and negative returns occur within the same realisation and offset one another over time. As $p$ increases, all three measures decline because heterogeneous neighbourhood signals suppress synchronised market-wide reversals and keep the aggregate opinion imbalance closer to zero.

Since $o_i(t)\in\{-1,0,1\}$, Eqs.~\eqref{eq:return_from_demand},
\eqref{eq:polarization}, and \eqref{eq:imbalance} imply

\begin{equation}
\label{eq:polarization_return_identity}
\mathcal P(t)=1-f_{\mathrm H}(t)-I(t)^2=1-f_{\mathrm H}(t)-\frac{r(t)^2}{\lambda^2}
\end{equation}

At fixed hold fraction, the inverse relation between return magnitude and opinion variance follows directly from the price-impact rule. The changes in polarisation in Fig.~\ref{fig:opinion_maps} should therefore be read together with the hold fraction and the absolute imbalance. The switching intensity provides additional information about the temporal dynamics.

\subsection{Strategy performance}

\begin{figure*}[tb]
    \centering
    \includegraphics[width=\textwidth]{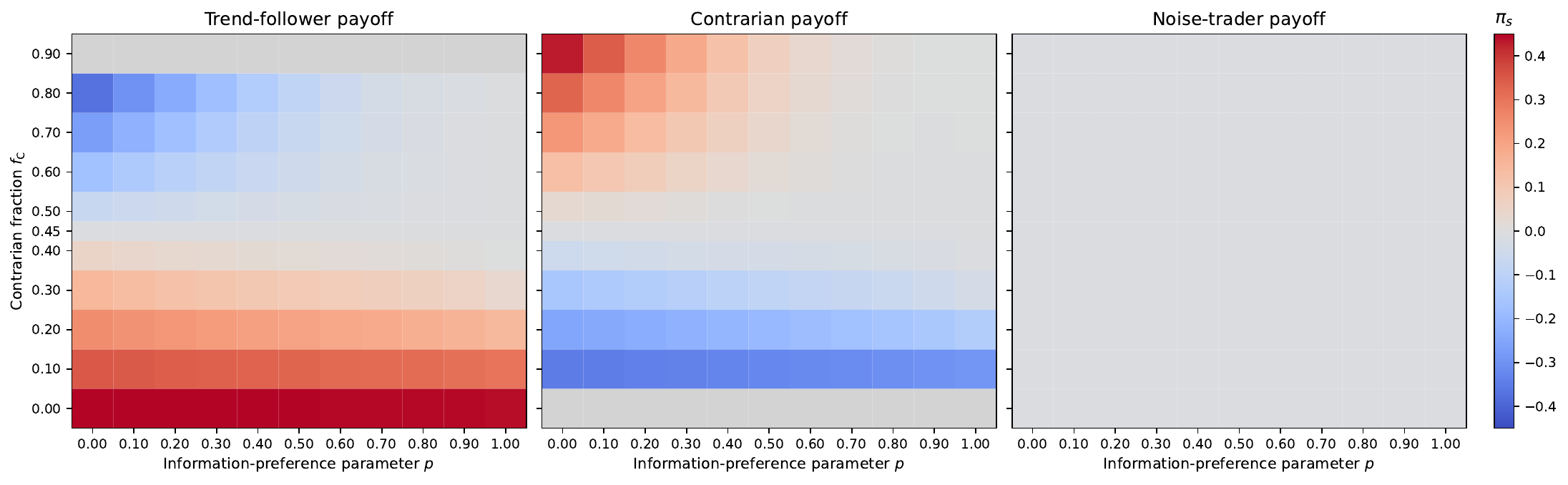}
    \caption{Mean per-agent payoff score per period as a function of the probability of selecting network information $p$ and the contrarian fraction $f_{\mathrm{C}}$ for trend followers (left), contrarians (middle), and noise traders (right). Payoffs are evaluated using the baseline switching cost $c=0.01$.}
    \label{fig:strategy_performance}
\end{figure*}

Figure~\ref{fig:strategy_performance} compares the mean payoff score defined in Eq.~\eqref{eq:payoff} for the three strategies. For $p=0$, $\theta=0$, and $c=0$, conditional on $r(t)\neq0$, the expected per-agent payoffs of trend followers and contrarians are given by

\begin{equation}
\label{eq:market_only_payoffs}
\begin{aligned}
\mathbb{E}\left[\pi_{\mathrm T}(t+1)\mid r(t)\neq0\right]
&=\lambda(f_{\mathrm T}-f_{\mathrm C}),\\
\mathbb{E}\left[\pi_{\mathrm C}(t+1)\mid r(t)\neq0\right]
&= -\lambda(f_{\mathrm T}-f_{\mathrm C}).
\end{aligned}
\end{equation}

Here, the expectation is taken over the noise traders' opinions. The strategy with the larger population fraction therefore has the higher expected payoff scores. At $p=0$, trend followers and contrarians also have equal switching costs after initialisation, so the cost does not affect their expected payoff difference.

When the contrarian fraction is low, trend followers obtain positive payoff scores, whereas contrarians obtain negative scores of similar magnitude. Increasing $p$ moderately reduces the absolute payoffs of both strategies, while the payoff of noise traders remains close to zero. This pattern is consistent with the persistent directional states (Fig.~\ref{fig:representative_regimes}). At intermediate contrarian fractions, the payoffs of trend followers and contrarians are much smaller. As $p$ increases, both approach zero and the payoff difference between the two strategies narrows. The dependence on the information source is strongest when the contrarian fraction is high. At low $p$, where agents rely mainly on the common market signal, contrarians obtain large positive payoff scores, whereas trend followers obtain large negative scores. As $p$ increases and local network information is selected more frequently, heterogeneous neighbourhood signals reduce the synchronisation of trading decisions. The payoffs of both strategies therefore decrease in magnitude and become closer. Noise traders obtain payoffs that are small compared with those of trend followers and contrarians. Appendix~\ref{app:strategy_performance} shows the dependence of the payoff differences on switching-cost assumptions. Across the range $c\in\{0,0.005,0.01,0.02\}$, the sign and broad parameter dependence of the payoff difference between contrarians and trend followers remain essentially unchanged.

\subsection{Network structure and node position}

\begin{figure*}[tb]
    \centering
    \includegraphics[width=\textwidth]{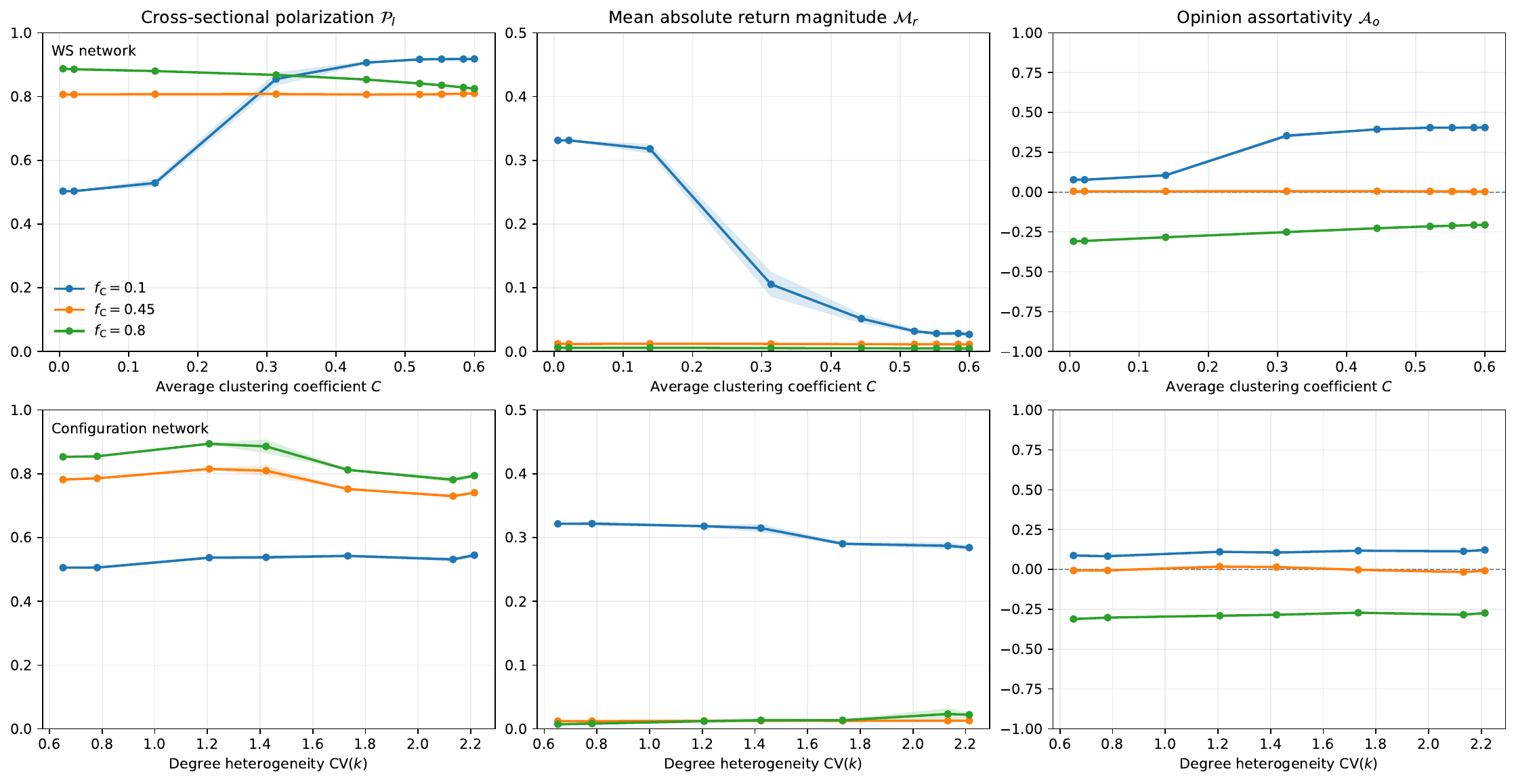}
    \caption{
    Collective opinion and market dynamics for different network structures under exclusive reliance on network information ($p=1$). The top row shows results for WS networks as functions of average clustering coefficient $C$, and the bottom row shows results for configuration networks as functions of degree heterogeneity $\mathrm{CV}(k)$. From left to right, the columns show cross-sectional polarisation $\mathcal P_I$, mean absolute return magnitude $\mathcal M_r$, and opinion assortativity $\mathcal A_o$. Each measure is calculated over the post-burn-in period and averaged across independent simulation runs. Shaded regions indicate approximate 95\% confidence intervals.}
    \label{fig:network_structure}
\end{figure*}

Figure~\ref{fig:network_structure} compares aggregate observables in WS and configuration networks. At $p=1$, returns do not feed back into the opinion updates. The strongest variation appears in WS networks when trend followers dominate ($f_C=0.1$). As average clustering increases, cross-sectional polarisation and opinion assortativity rise markedly, whereas the mean absolute return magnitude declines. The higher assortativity indicates greater local opinion alignment, while the increase in polarisation indicates greater disagreement across the population. Clustering can also facilitate the formation of locally homogeneous opinion groups, although its effects on population-level polarisation depend on information reinforcement and individual stubbornness~\cite{Botte2022}. The smaller aggregate imbalance gives smaller return magnitudes through $r(t)=\lambda I(t)$. Changing the WS rewiring probability also affects other graph properties, including path length, so these trends cannot be attributed to clustering alone. When the strategy composition is more balanced ($f_C=0.45$), all three aggregate measures are comparatively insensitive to clustering. When contrarians dominate ($f_C=0.8$), returns also remain small; increasing clustering produces only a moderate reduction in polarisation and makes the negative opinion assortativity less pronounced. Thus, the dependence on WS rewiring is strongest when trend followers dominate. In contrast, varying degree heterogeneity in the configuration networks has a limited effect on the aggregate dynamics. Polarisation, return magnitude, and opinion assortativity change only modestly across the observed range of $\mathrm{CV}(k)$.

\begin{figure*}[tb]
    \centering
    \includegraphics[width=\textwidth]{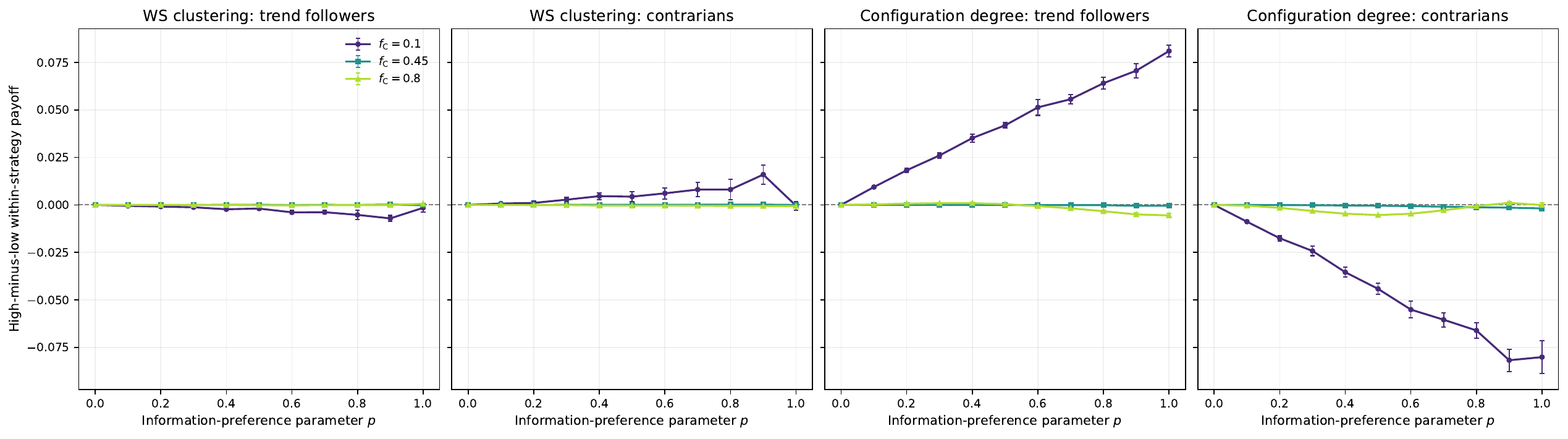}
    \caption{Differences in the payoff scores across network positions within each strategy. The panels show high-minus-low payoff differences between the top and bottom quartiles of local clustering in WS networks and node degree in configuration networks. Positive values indicate higher payoff scores for agents in the upper quartile. Each measure is calculated over the post-burn-in period and averaged across independent simulation runs. Error bars denote approximate 95\% confidence intervals.}
    \label{fig:node_position}
\end{figure*}

Figure~\ref{fig:node_position} shows when the aggregate structural effects are accompanied by systematic payoff differences across node positions. The high-minus-low clustering payoff-score differences remain close to zero for most values of $p$ and $f_C$, and the deviations at high $p$ are small relative to their uncertainty. Thus, the changes in the collective observables across WS networks are not accompanied by a consistent payoff-score difference between the local-clustering quartiles. Node degree shows a clearer association with payoff scores when trend followers dominate. For $f_C=0.1$, the payoff premium of high-degree trend followers becomes increasingly positive as $p$ rises, whereas the corresponding difference for contrarians becomes negative. One possible explanation is that high-degree agents sample more neighbours, so their network signal may agree more often with the population majority when trend followers dominate. Under the payoff rule, such alignment tends to increase trend-follower scores and decrease contrarian scores. These score differences do not imply more accurate information or realised trading profit. The degree-based differences are much weaker when the strategy composition is balanced or when contrarians dominate. Aggregate observables thus change little across the degree distributions studied, although payoff scores differ between high- and low-degree agents.

\section{Conclusions}
\label{conclusions}

Separating information-source selection from fixed response strategies reveals three dynamical regimes as the probability of selecting the network signal and the fraction of contrarians vary. For $p=0$ and $\theta=0$, the deterministic limit gives a threshold $f_{\mathrm{C}}^*=(1-f_{\mathrm{N}})/2$, separating persistent from alternating return signs. When trend followers dominate, early stochastic fluctuations select a market direction that is subsequently reinforced, producing persistent directional opinions and returns. When contrarians dominate and investors rely primarily on the common market signal, their synchronised response produces frequent reversals between buy- and sell-dominated states, together with high volatility and large positive and negative returns. When the same contrarian-rich population relies mainly on heterogeneous network information, opposing buy and sell opinions coexist within each cross section and largely offset one another, yielding comparatively small aggregate returns. This coexistence of opposing opinions does not necessarily imply their segregation into distinct opinion clusters.

In Watts–Strogatz networks at $p=1$, higher clustering along the rewiring series is associated with greater opinion assortativity and polarisation and smaller mean absolute returns when trend followers dominate. Degree heterogeneity in configuration networks has a comparatively limited effect on aggregate dynamics. Nevertheless, payoff differences between high- and low-degree agents are observed. When trend followers dominate and network information plays an important role, high-degree trend followers have higher payoff scores than low-degree trend followers, whereas high-degree contrarians have lower scores than low-degree contrarians. Under common market information, the more numerous of the two directional strategies has the higher expected payoff scores, consistent with the contemporaneous price-impact rule.

Investment strategies and network structure remain fixed, opinions are restricted to three states, and the price-impact mechanism does not represent an order book, explicit inventory accounting, or a fundamental asset value. The reported payoff scores measure alignment with contemporaneous model returns rather than realised trading profits. In contrarian-dominated populations, local information weakens the synchronised reversals induced by common market feedback, without eliminating disagreement among agents.

\begin{acknowledgments}
Y.Y. was supported by the China Scholarship Council (grant no. 202508310156). L.E.C.R. was partially supported by the Belgian Special Research Fund (BOF) (2024/01/709).
\end{acknowledgments}

\appendix

\section{Opinion and return distributions for representative regimes}
\label{app:representative_regimes}

\subsection{Opinion imbalance and cross-sectional composition}

\begin{figure*}[tb]
    \centering
    \includegraphics[width=\textwidth]{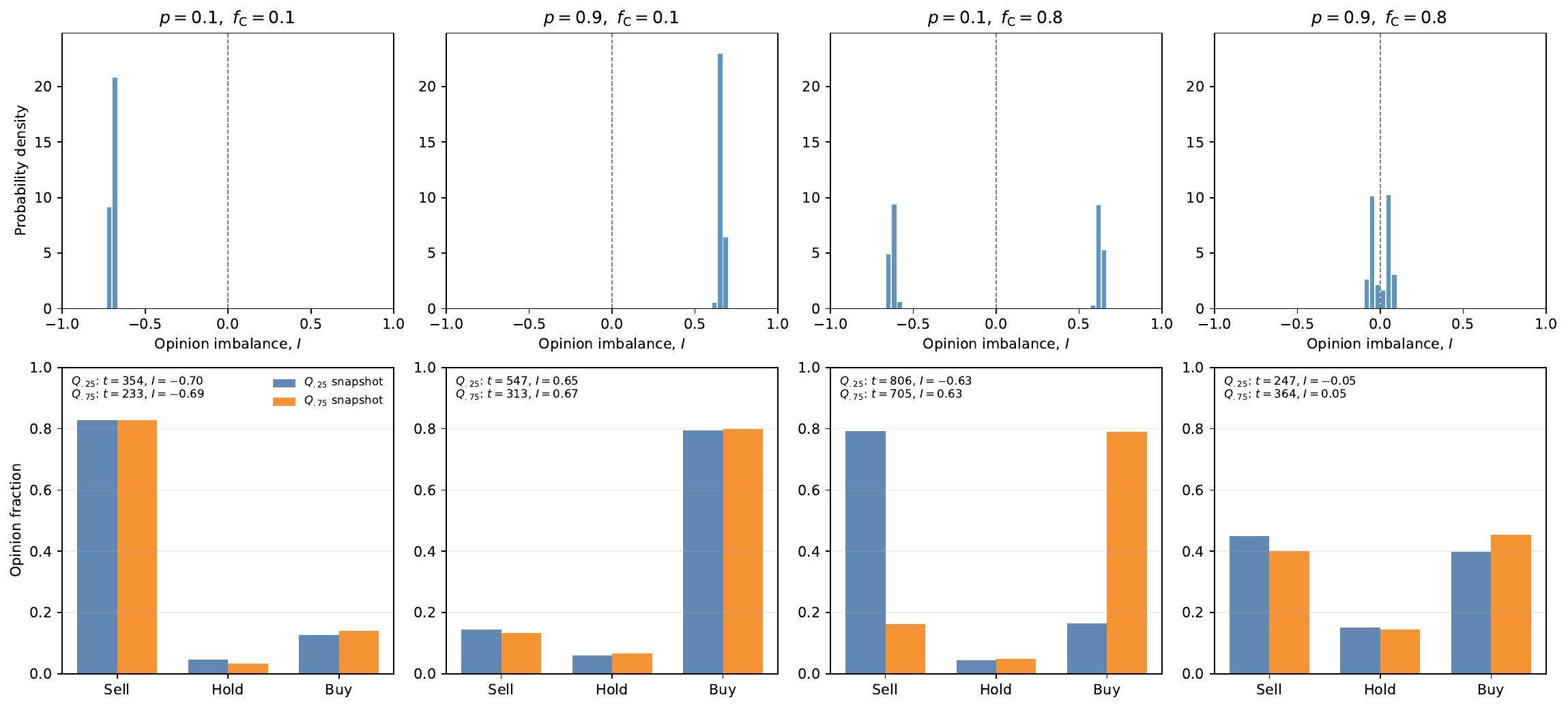}
    \caption{Imbalance distributions and opinion fractions for the representative opinion regimes. From left to right, the columns correspond to $(p,f_{\mathrm{C}})=(0.1,0.1)$, $(0.9,0.1)$, $(0.1,0.8)$, and $(0.9,0.8)$. The upper row shows the within-run probability density of the aggregate opinion imbalance $I(t)$ over the post-burn-in period; the vertical dashed line indicates $I=0$. The lower row shows the fractions of sell, hold, and buy opinions at two time points selected as the observations closest to the 25th and 75th percentiles of the post-burn-in imbalance.}
    \label{fig:opinion_distribution}
\end{figure*}

For each representative realisation, the upper row of Fig.~\ref{fig:opinion_distribution} shows the distribution of the aggregate opinion imbalance, $I(t)=f_{\mathrm{B}}(t)-f_{\mathrm{S}}(t)$, over the post-burn-in period. At the low contrarian fraction, $f_{\mathrm{C}}=0.1$, the imbalance distribution in both representative realisations is unimodal and concentrated on one side of zero. The two cross-sectional snapshots within each realisation also have similar compositions. Sell opinions remain dominant at $p=0.1$, whereas buy opinions remain dominant at $p=0.9$. These patterns show that both realisations are persistent directional-majority states. Their opposite directions reflect stochastic directional selection rather than a systematic effect of $p$.

A different pattern appears at $f_{\mathrm{C}}=0.8$ and $p=0.1$. The within-run imbalance distribution has two modes with opposite signs, and the two selected snapshots are dominated by opposite directional opinions. The bimodality therefore reflects repeated temporal switching between buy- and sell-dominated states, rather than the simultaneous coexistence of balanced buy and sell fractions within a single cross section.

At $f_{\mathrm{C}}=0.8$ and $p=0.9$, the imbalance distribution is instead unimodal and centred near zero. Both cross-sectional snapshots contain substantial and approximately balanced fractions of buy and sell opinions. Because the hold fraction remains comparatively small, the near-zero imbalance primarily reflects the simultaneous coexistence of opposing directional opinions rather than widespread holding.

\subsection{Return distributions across independent simulation runs}

\begin{figure*}[tb]
    \centering
    \includegraphics[width=\textwidth]{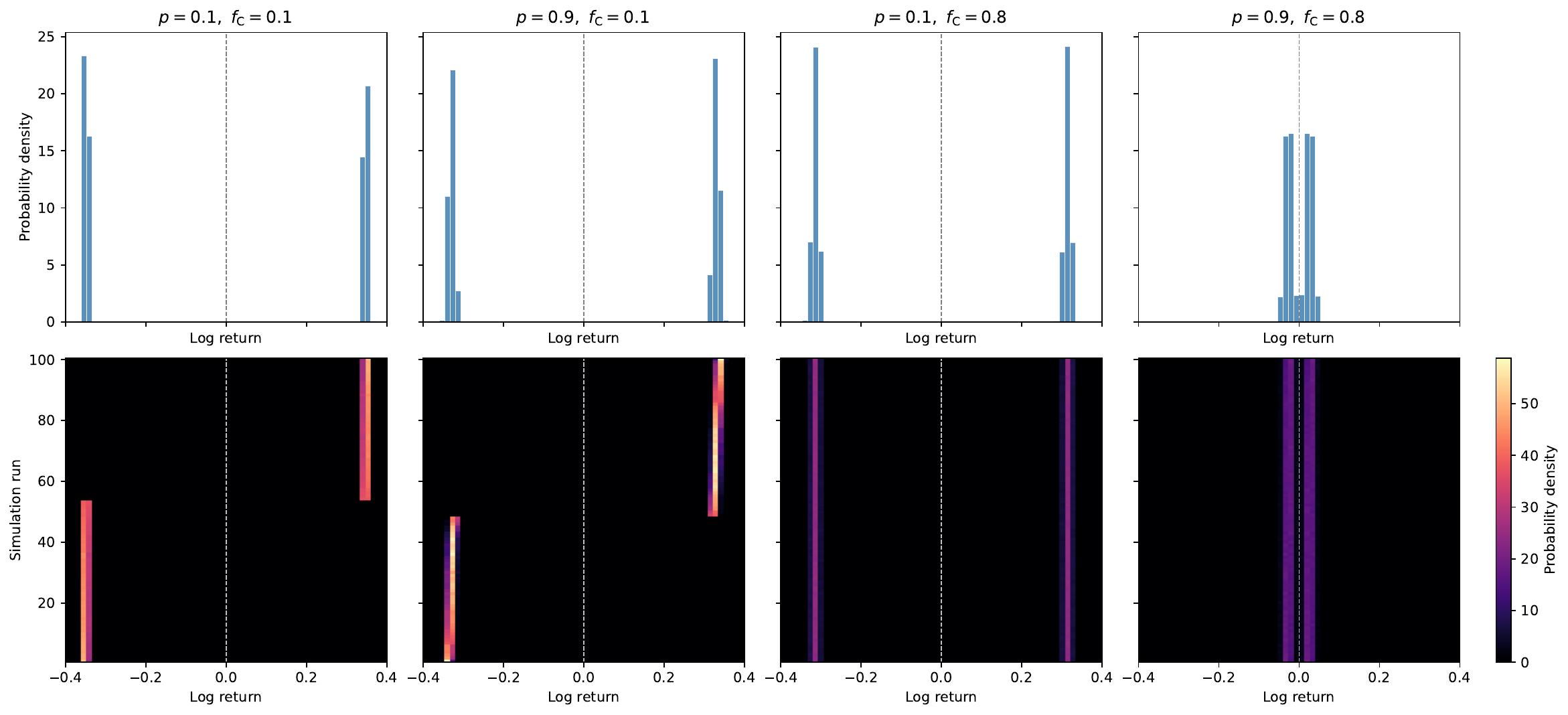}
    \caption{Return distributions across simulation runs. From left to right, the columns correspond to $(p,f_{\mathrm{C}})=(0.1,0.1)$, $(0.9,0.1)$, $(0.1,0.8)$, and $(0.9,0.8)$. The upper row shows the probability density of post-burn-in log returns pooled across all runs. The lower row shows the corresponding run-specific probability densities, with runs ordered by their mean return. The vertical dashed line indicates zero return, and the colour scale represents probability density.}
    \label{fig:return_distributions}
\end{figure*}

\begin{figure*}[tb]
    \centering
    \includegraphics[width=\textwidth]{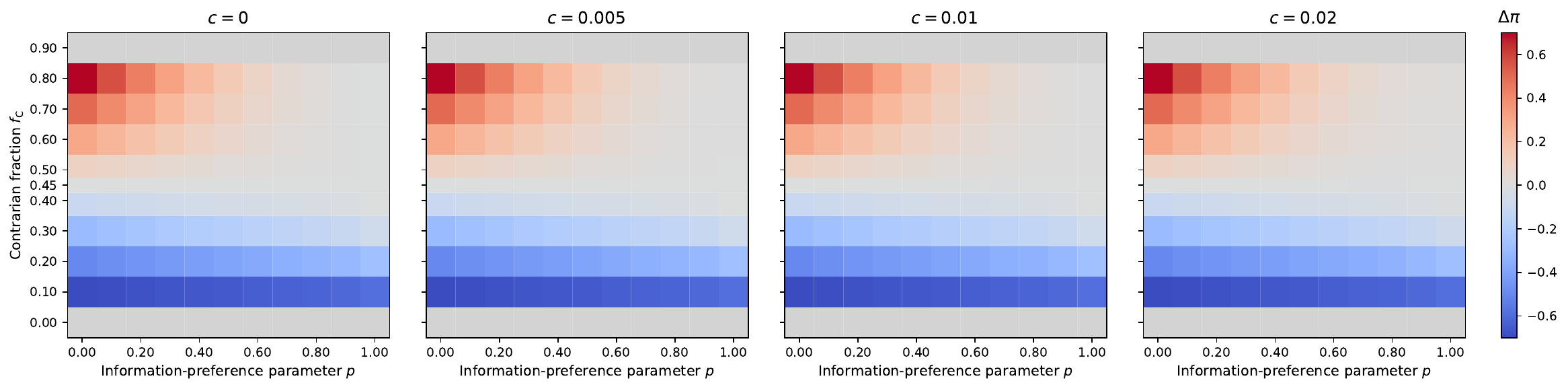}
    \caption{Sensitivity of relative strategy payoff scores to switching costs. Each panel shows the payoff difference $\Delta\pi=\pi_C-\pi_T$ across the information-preference parameter $p$ and the contrarian fraction $f_C$ for a different switching cost $c$. Positive values indicate a higher mean payoff score for contrarians, whereas negative values indicate a higher score for trend followers.}
    \label{fig:strategy_cost_robustness}
\end{figure*}

Figure~\ref{fig:return_distributions} compares the return distributions across repeated simulation runs for the same four parameter combinations. The pooled distributions in the upper row show the overall range of outcomes across runs, whereas the run-specific distributions in the lower row reveal whether different return modes occur across separate realisations or within the same realisation.

When the contrarian fraction is low, the pooled return distributions contain both negative and positive modes. The run-specific distributions, however, show that each realisation is generally concentrated around only one of these modes. This pattern occurs at both low and high values of $p$, consistent with the buy-sell symmetry of the model. Instead, early stochastic fluctuations lead each realisation toward either a persistent buy-dominated state with positive returns or a persistent sell-dominated state with negative returns. When the contrarian fraction is high and $p=0.1$, both negative and positive return modes occur within the same realisation. Reliance on a common market signal coordinates the response of the contrarian majority, repeatedly reversing the aggregate opinion imbalance and producing alternating returns. By contrast, when $f_{\mathrm{C}}=0.8$ and $p=0.9$, both the pooled and run-specific return distributions are concentrated near zero. Heterogeneous neighbourhood signals weaken synchronised market-wide responses, so opposing local buying and selling opinions largely offset one another in aggregate.

\subsection{Sensitivity of payoff scores to switching costs}
\label{app:strategy_performance}

To assess whether the relative payoff scores of trend followers and contrarians depend on the assumed switching cost, we recompute strategy payoffs for $c\in\{0,0.005,0.01,0.02\}$. Since $c$ enters only the payoff definition, changing it does not alter the opinion or return trajectories. For each parameter combination, payoffs are first averaged across repeated simulations within each network realisation, after which the paired payoff difference is averaged across networks.

Figure~\ref{fig:strategy_cost_robustness} shows that the sign and parameter dependence of the payoff difference change little over the four switching-cost values. At low contrarian fractions, $\Delta\pi$ remains negative, indicating that trend followers have higher payoff scores. At high contrarian fractions, $\Delta\pi$ remains positive, indicating higher scores for contrarians. In both cases, the magnitude of the payoff difference generally decreases as $p$ increases, consistent with the weakening of market-wide coordination when investors rely more heavily on heterogeneous local information. Around the balanced strategy composition, the payoff difference remains close to zero.

\bibliographystyle{apsrev4-2}
\bibliography{references}

\end{document}